\documentclass[prd,aps,10pt,twocolumn,showpacs,reprint,nofootinbib]{revtex4}
 
\usepackage[dvips]{graphicx}
\usepackage{amssymb}
\usepackage{filecontents}
\usepackage{amsmath}
\usepackage{color}
\begin{document}
\title{Compact stars in Energy-Momentum Squared Gravity}
\author{Nasrin Nari}
\author{Mahmood Roshan\footnote{mroshan@um.ac.ir}}

\affiliation{Department of Physics, Ferdowsi University of Mashhad, P.O. Box 1436, Mashhad, Iran}

\begin{abstract}
A simple generalization to Einstein's general relativity (GR) was recently proposed which allows a correction term $T_{\alpha\beta}T^{\alpha\beta}$ in the action functional of the theory. This theory is called Energy-Momentum Squared Gravity (EMSG) and introduces a new coupling parameter $\eta$. EMSG resolves the big bang singularity and has a viable sequence of cosmological epochs in its thermal history. Interestingly, in the vacuum EMSG is equivalent to GR, and its effects appear only inside the matter-energy distribution. More specifically, its consequences appear in high curvature regime. Therefore it is natural to expect deviations form GR inside compact stars. In order to study spherically symmetric compact stars in EMSG, we find the relativistic governing equations. More specifically, we find the generalized version of the Tolman-Oppenheimer-Volkov equation in EMSG. Finally we present two analytical solutions, and two numerical solutions for the field equations. For obtaining the numerical solutions we use polytropic equation of state which is widely used to understand the internal structure of neutron stars in the literature. Eventually we find a mass-radius relation for neutron stars. Also, We found that EMSG, depending on the central pressure of the star and the magnitude of free parameter $\eta$, can lead to larger or smaller masses for neutron stars compared with GR. Existence of high-mass neutron stars with ordinary polytropic equation of state in EMSG is important in the sense that these stars exist in GR when equation of state is more complicated. 
\end{abstract}
\maketitle
\section{introduction}
Einstein's general relativity (GR) is the most successful theory of gravity and can explain a wide variety of gravitational phenomena from local to large structure in the universe. Specifically, after decades of being under intense scrutiny, it is  well-established that GR passes the local solar system tests successfully. On the other hand, at cosmological scales, the standard cosmological model, i.e. the $\Lambda$CDM model, based on GR is the most complete model to explain the dynamics of the cosmos. {More importantly, recent observations proved that gravitational wave, as one the main predictions of GR, exist and its power spectrum and properties are consistent with GR's descriptions \cite{gwaves}.}

 {However, there are several unresolved issues which keep open the way to frameworks which try to extend GR. For some examples of unresolved problems in GR we mention the dark matter problem at the galactic and cosmological scales, the dark energy enigma, and the presence of singularities in the early universe and inside black holes. It is interesting that for all of these problems, modifying GR can help to find a solution. For instance, for the dark matter problem, there are some modified theories of gravity which are relatively successful to explain the dark matter observations. For widely studied alternative theories to dark matter particles, we mention the modified-Newtonian-dynamics (MOND) \cite{mil} and its relativistic formulation TeVeS \cite{beck}, and scalar-tensor-vector gravity theory known as MOG \cite{moff}. On the other hand, there are several dark energy models which generalize GR to explain the cosmic speed up without the cosmological constant $\Lambda$. {We remind that, existence of $\Lambda$ in GR's generic action causes the cosmological constant problem, which can be considered as a serious inconsistency between GR and quantum field theory.} For a comprehensive review on the subject we refer the reader to \cite{capo1}.} 
 
{Existence of singularities {in GR can be viewed as a problem in the sense that GR predicts it at high energy regimes where GR itself is no longer valid because of the expected quantum effects.} However we know that there is no precise formulation for quantum gravity. Therefore, there are some classical models in which the big bang singularity can be resolved in non-quantum approaches. For one of the recent theories we refer to Eddington-inspired Born-Infeld (EiBI) theory \cite{EiBI}. This theory is equivalent to GR in vacuum and its effects appear only inside matter sources. For other recent attempts to resolve the cosmological singularities by using extensions of GR, see \cite{sing} and references therein.} 
 
{Recently another covariant generalization to GR has been presented in \cite{kat,us1}. This theory allows an specific coupling between matter and gravity. More specifically, GR has been modified by adding a nonlinear $T_{\alpha\beta}T^{\alpha\beta}$ term to the generic action, where $T_{\alpha\beta}$ is the energy-momentum tensor. The authors in \cite{us1} called this theory as energy-momentum-squared gravity (EMSG). Although EMSG seems simple, leads to several interesting consequences. For example this theory has a regular bounce in the early universe and possesses a minimum length scale, and a finite maximum energy density at early universe. Therefore EMSG can resolve big bang singularity with a classic and non-quantum prescription. For a detailed study of the model and its cosmological consequences, we refer the reader to \cite{us1}. In \cite{bar} a general version of EMSG has been investigated and several interesting exact cosmological solutions have been found.}

{Since EMSG is introduced to resolve the singularities, it is natural to expect its deviations from GR appear only at high energy/curvature regimes. Therefore it is also necessary to investigate it inside compact stars where energy scale is high enough to see EMSG deviations from GR}. In this paper we study relativistic stars in the context of EMSG. In other words, we are interested in spherically symmetric and static solutions of EMSG in the presence of matter. It is necessary to emphasize again that EMSG is equivalent to GR in vacuum. 

We find the governing equations including the modified Tolman-Oppenheimer-Volkov equation in EMSG and solve them for polytropic models of neutron stars. We also find a mass-radius relationship for neutron stars, and show that EMSG supports high-mass neutron stars. It should be stressed that recent discovery of high-mass neutron stars \cite{ns} rules out many standard equations of state (EOS) in GR. A great effort is still underway to estimate approperiate EOS for neutron stars, for example see \cite{mosh}. The other possibility, which has been widely investigated is to interpret the observations in terms of modified gravity effects at large curvature, for some recent works which follow this direction see \cite{ns1}. From this perspective, neutron stars seem as appropriate lab to test modified gravity theories.

The outline of the paper is the following. In section \ref{1} we briefly introduce EMSG and its field equations. In section \ref{mo} we introduce the main equations governing the spherically symmetric matter distribution. In section \ref{analytic} we find two exact solutions for the field equations of EMSG. Furthermore, in section \ref{numeric} we study polytropic and strange quark stars numerically. Finally, we summarize the results in the Conclusion section.

\section{Brief introduction to energy-momentum squared gravity}
\label{1}
{Let us start with the action of EMSG presented in \cite{us1}}
\begin{equation}
S=\frac{1}{2\kappa}\int \sqrt{-g}\left(R-2\Lambda-\eta \mathbf{T}^2\right)d^4x+S_M
\label{action}
\end{equation}
where $R$ is the Ricci scalar, $\kappa=8\pi G/c^4$, $\Lambda$ is the cosmological constant, and $S_M$ is the matter action defined as 
\begin{equation}
S_{M}=\int L_m \sqrt{-g}d^4x
\end{equation}
in which $L_m$ is the matter Lagrangian density. The energy-momentum tensor then is usually defined as follows
\begin{equation}
T_{\mu\nu}=-\frac{2}{\sqrt{-g}}\frac{\delta (\sqrt{-g}L_m)}{\delta g^{\mu\nu}}
\end{equation}
Finally, $\mathbf{T}^2$ in the action is given by $\mathbf{T}^2=T_{\alpha\beta}T^{\alpha\beta}$.

Appearance of $T_{\mu\nu}$ in the gravitational action may arise an important question as: How can matter know the distribution of itself in advance, i.e. before the action is varied? To clarify this question let us conveniently assume that the action of the theory is written as $S=S_G+S_M$, where $S_G$ is the gravitational action. To find the ordinary matter energy-momentum tensor, it is not necessary to know anything about $S_G$. Therefore irrespective of the gravitational theory, one can vary $S_m$ (and not $S_G$) with respect to the metric tensor and, in principle, find $T_{\mu\nu}$ in terms of physical variables for a given matter source. For example these physical variables for a perfect fluid are the velocity, the density and the pressure of the fluid\footnote{Of course determining the energy momentum tensor, does not mean that we know the mass/energy distribution in terms of $x^{\alpha}$. More specifically, one still needs dynamical differential equations obtained from variation of $S$ with respect to the matter fields and the metric tensor to find the physical functions in terms of space-time coordinates $x^{\alpha}$.}. In fact one only needs the matter Lagrangian density $L_m$. In this case $T_{\mu\nu}$ can be written in terms of $L_m$, $g_{\mu\nu}$ and derivatives of $L_m$ with respect to $g_{\mu\nu}$, for more details see \citep{Harko}. From this perspective, in principle, one can assign a well-defined energy-momentum tensor to any physical mass/energy source. Now the above mentioned question can be asked in a different way: Are we allowed to use $T_{\mu\nu}$ to construct scalars to be included in the gravitational action $S_G$? Or equivalently can we use $L_m$ and its derivatives, like $\partial L_m/\partial g_{\mu\nu}$ to construct scalars to be included in $S_G$?

Although including such scalars in the gravitational action causes new couplings between matter and gravity which are absent in GR, there is not a-priori fundamental reason to prevent them. Consequently, this kind of modifications has received a continuous interest in the last decade. For example for theories which introduce $T_{\mu\nu}$ in the action functional of the theory, we refer the reader to \citep{Harko,Haghani:2013oma}, where correction terms including $g_{\mu\nu} T^{\mu\nu} $ and $R_{\mu\nu}T^{\mu\nu}$ appear in the action respectively. For a model in which $L_m$ (and not its derivatives) linearly appears in $S_G$ see \citep{bertolami,faraoni}. 

 This generalization to GR, i.e. \eqref{action}, has been also investigated in \cite{gcoupling}. In this paper we assume the metric signature as $(-,+,+,+)$. Also $\eta$ is a coupling constant and its magnitude can be constrained by observations. In fact EMSG is a one parameter theory and introduced only one free parameter. It has been shown in \cite{us1} that $\eta>0$ otherwise it leads to wrong cosmological epochs. More specifically, there is no stable de Sitter universe when $\eta<0$. Therefore, in this paper we will focus on EMSG with $\eta>0$. {One may simply find crude constraint on $\eta$. For example, considering the correction term in the action (\ref{action}), in the current phase of the universe we expect that $\eta \mathbf{T}^2 \ll \Lambda$. Equivalently, EMSG's effects can be ignored in the dark energy dominated phase of the cosmos if}
\begin{equation}
\eta\ll \frac{\Lambda}{\rho_{\text{0,crit}}^2 c^4}\simeq 10^{-17}\text{s}^4\, \text{kg}^{-2}
\label{et1}
\end{equation}
where $\rho_{\text{0,crit}}\simeq 10^{-25}\text{kg}\,\text{m}^{-3}$ is the critical present density, and $\Lambda\simeq 10^{-52}\text{m}^{-2}$. Furthermore, it has been shown in \cite{us1} that EMSG could resolve the big bang singularity in a non-quantum way, provided that the maximum density and minimum length scale, which appear at the early universe of EMSG, are smaller than the Planck density and length respectively. In this case, it is necessary to assume 
\begin{equation}
\eta>\Big(\frac{\kappa}{8\pi}\Big)^3\, \hbar c\simeq 10^{-158}\text{s}^4\, \text{kg}^{-2}
\label{et2}
\end{equation}
where $\hbar$ is the reduced Planck constant. Recently\footnote{We need to mention that when we were preparing this paper for submission, the paper \cite{aka2} appeared on arXiv. Regarding the existence of some similarities between our paper and \cite{aka2}, we also sunsequently posted our paper on arXiv. }, parameter $\eta$ has been constrained using observational measurements of neutron stars in \cite{aka2}. Their bound on positive $\eta$ can be written as $\eta< 4.14\times 10^{-80}\,\text{s}^4\, \text{kg}^{-2}$, which is consistent with conditions \eqref{et1} and \eqref{et2}.

{It is clear that in vacuum EMSG is equivalent to GR, and inside matter sources is different and, in principle, can lead to new consequences.} In fact although Einstein's field equations involve the energy-momentum tensor linearly, in EMSG there are nonlinear corrections constructed by $T_{\mu\nu}$. {In EMSG, the metric tensor is the only field attributed to gravity, and one can find the corresponding field equations by} varying the action \eqref{action} with respect to metric tensor \cite{us1}. The result can be written as   
\begin{equation}
G_{\mu\nu}+\Lambda g_{\mu\nu}=\kappa T_{\mu\nu}^{\text{eff}}
\label{fe1}
\end{equation}
{where an effective energy-momentum tensor is defined as}
\begin{equation}
T_{\mu\nu}^{\text{eff}}=T_{\mu\nu}+2\frac{\eta}{\kappa}\left(\mathbf{\Psi_{\mu\nu}}+T_{\mu}^{\sigma}T_{\nu\sigma}-\frac{1}{4}g_{\mu\nu}\mathbf{T}^2\right)
\label{n11}
\end{equation}
in which, for a perfect fluid, $\mathbf{\Psi_{\mu\nu}}$ is expressed as
\begin{equation}
\mathbf{\Psi_{\mu\nu}}=-L_m S_{\mu\nu}-\frac{1}{2}T T_{\mu\nu}-2T^{\alpha\beta}\frac{\partial^2 L_m}{\partial g^{\alpha\beta}\partial g^{\mu\nu}}
\label{n4}
\end{equation}
where $S_{\mu\nu}=T_{\mu\nu}-T g_{\mu\nu}/2$, $T$ is the trace of the energy-momentum tensor. For a perfect fluid system, one may simply define $L_{M}$ as $L_M=p$, for more details see \cite{barrow}. {It is straightforward to show that the effective energy momentum tensor is conserved, i.e. $\nabla^{\mu}T_{\mu\nu}^{\text{eff}}=0$. {This fact directly means that the ordinary matter energy-momentum tensor is not conserved. Naturally this fact puts constraint on the parameter $\eta$ to make EMSG consistent with the current observations. }

\section{Relativistic compact stars in EMSG: Generalized Tolman-Oppenheimer-Volkov equation}
\label{mo}
{In order to find the governing equations describing the internal structure of a} star in EMSG, let us start with the general form of the static and spherically symmetric metric
\begin{equation}
ds^2=-b(r)c^2 dt^2+\frac{dr^2}{f(r)}+r^2d\Omega^2
\label{me}
\end{equation}
{Throughout this work we assume a perfect fluid energy-momentum tensor for the compact object as}
\begin{equation}
T_{\mu\nu}=(\rho+\frac{p}{c^2})u_{\mu}u_{\nu}+p\, g_{\mu\nu}
\end{equation}
{where $p$, $\rho$ and $u^{\mu}$ are the pressure, energy density, and four velocity of the fluid, respectively. We mention again that outside matter fields, EMSG coincides with GR. In \cite{us1} an exact spherically symmetric solution outside a charged black hole, where the matter fields do not vanish, has been found. {Using the above mentioned metric and $T_{\mu\nu}$, it} is straightforward to show that the "tt" and "rr" components of the field equation \eqref{fe1} can be written as } 
\begin{equation}
rf'+f-1+{\kappa} c^2 {\rho_{\text{eff}}(r)}\,r^2=0
\label{two}
\end{equation}
\begin{equation}
f\,b'\,r+b\left(f-1\right)-{\kappa}\,p_{\text{eff}}(r)\,b\,r^2=0
\label{one}
\end{equation}
where the prime stands for derivative with respect to $r$, and $\rho_{\text{eff}}(r)$ and $p_{\text{eff}}(r)$ are defined as
\begin{equation}\label{efro}
\rho_{\text{eff}}(r)=\rho(r)+\frac{\Lambda}{\kappa c^2}-\frac{\eta c^2}{2 \kappa}\left(8{\rho}\,\frac{p}{c^2}+\,\rho^2+3 \,\frac{p^2}{c^4}\right)
\end{equation}
\begin{equation}\label{efp}
p_{\text{eff}}(r)=p(r)-\frac{\Lambda}{\kappa}-\frac{\eta c^4 }{2\kappa}\Big(\rho^2+3\,\frac{p^2}{c^4}\Big)
\end{equation}

Hereafter we assume that the cosmological constant is zero. {This assumption is not restrictive in the sense that} it is natural to expect that $\Lambda$ has no effect inside massive stars.

{ On the other hand instead of using the component "$\theta\theta$" let us use the equation obtained from the consevation of effective energy-momentum tensor, i.e. $\nabla^{\mu}T_{\mu\nu}^{\text{eff}}=0$. The only non-zero component of this equation is given by} 
\begin{equation}
b'\left(\rho_{\text{eff}} + \frac{p_{\text{eff}}}{c^2}\right)+2 b \frac{p'_{\text{eff}}}{c^2} =0
\label{three}
\end{equation}
{It is clear that be setting $\eta$ to zero, the standard equations in GR are recovered. Equations \eqref{two}, \eqref{one} and \eqref{three} are three differential equations for four unknown functions $b(r)$, $f(r)$, $\rho(r)$ and $p(r)$. Therefore we need an EOS in order to construct a complete set of differential equations. Before moving on to discuss the boundary conditions, let us combine equations \eqref{two}, \eqref{one} and \eqref{three} in order tot find the generalized version of the Tolman-Oppenheimer-Volkov (TOV) equation. It is straightforward to integrate \eqref{two} and write $f(r)$ as}
\begin{equation}
f(r)=1-\frac{\kappa c^2 \,m_{\text{eff}}(r)}{4\pi r}
\label{effmass}
\end{equation}
{where $m_{\text{eff}}$ is an effective mass parameter defined as 
\begin{equation}
m_{\text{eff}}(r)=4\pi\int_0^r \rho_{\text{eff}}(r) r^2 dr
\label{emass}
\end{equation}
Using equations \eqref{one} and \eqref{effmass}, the conservation equation \eqref{three} can be rewritten as follows}

\begin{equation}\label{tolman}
\begin{split}
\frac{d p_{\text{eff}}}{dr}=&-\frac{\rho_{\text{eff}}c^2+p_{\text{eff}}}{r^2}\Big(\frac{\kappa}{2}p_{\text{eff}}r^3+\frac{\kappa c^2 m_{\text{eff}}(r)}{8\pi}\Big)\times\\&~~~~\Big(1-\frac{\kappa c^2 m_{\text{eff}}(r)}{4\pi r}\Big)^{-1}
\end{split}
\end{equation}
Naturally by setting the parameter $\eta$ to zero, the standard TOV equation is recovered.

It is interesting that the EMSG corrections appear in the field equations as effective energy density and pressure. In other words the main equations \eqref{two}, \eqref{one}, \eqref{three}, and TOV equation \eqref{tolman} are exactly the same as the corresponding equations in GR. The only difference is that the effective quantities $\rho_{\text{eff}}$ and $p_{\text{eff}}$ appear instead of $\rho$ and $p$. This point induces a remarkable mathematical simplicity in our analysis. 

From equations \eqref{efro} and \eqref{efp}, at least at first glance, one may infer that EMSG corrections, effectively, reduce the energy density and pressure in the system. Therefore, one may conclude that EMSG weakens the gravitational strength caused by a given $\rho$ and $p$. This behavior is totally in agreement with the main aim of the theory which is preventing the singularities. It is necessary to emphasize that at relativistic situations, the pressure has gravitational effects and, in principle, can support the local gravitational collapse instead of preventing it. For example, in the Post-Newtonian regime where the relativistic effects are important, it has been shown that by increasing the pressure, the Jeans mass decreases \cite{nazar}. This means that pressure can trigger the gravitational instability. 

However although it seems satisfactory, it must be stressed that reducing the pressure effectively, does not necessarily mean that the gravitational strength has been weaken. In other words, in non-relativistic situations, reducing the pressure supports the gravitational collapse. In other words, in this case reduction in pressure highlights the gravitational effects. Therefore, as we will see in the next section, the overall influence of EMSG on the mean density and other properties of the stars is not trivial. In other words in some cases EMSG lead to more compact stars compared to GR, and in some other cases leads to less compact stars. 

{In the subsequent sections we solve the governing equations using analytic and numeric prescriptions.}
\section{Analytic solutions}
\label{analytic}
Keeping in mind the mathematical similarity between EMSG equations and GR, one may expect that all analytic solutions for $\rho$ and $p$ in GR are also valid in EMSG for $\rho_{\text{eff}}$ and $p_{\text{eff}}$. In the following let us consider some special cases, and present two exact solutions for the governing equation.
\subsection{Case $\rho_{\text{eff}}=constant$}
One of the well known analytic solutions in GR is the Schwarzschild constant-density interior solution. Therefore, it is straightforward to verify that when the effective density is constant inside radius $R$ (i.e. $\rho_{\text{eff}}=\rho_0$), and is zero for $r>R$, the effective pressure inside the star is given by
\begin{equation}
p_{\text{eff}}(r)=\rho_{0}c^2\frac{\sqrt{1-\frac{\kappa c^2 m_{\text{eff}} }{4\pi R^3}r^2}-\sqrt{1-\frac{\kappa c^2 m_{\text{eff}}}{4\pi R}}}{3 \sqrt{1-\frac{\kappa c^2 m_{\text{eff}}}{4\pi R}}-\sqrt{1-\frac{\kappa c^2 m_{\text{eff}} }{4\pi R^3}r^2}}
\label{ints}
\end{equation}
where $m_{\text{eff}}$ is an constant mass obtained from \eqref{emass}. In fact $m_{\text{eff}}$ is the mass parameter of the star which appear in the external Schwarzschild external solution. {It is interesting that although the mathematical form of the metric outside a spherical energy-matter distribution is the same in EMSG and GR, their mass parameters are different}. 

The solution (\ref{ints}) represents a star in EMSG with a complicated equation of state given by
\begin{equation}
p=c^2\sqrt{\frac{2\kappa}{3\eta}(\rho-\rho_0)+\frac{13}{9}\rho^2}-\frac{4}{3}\rho c^2
\label{eo1}
\end{equation}
For the corresponding solution in GR, the density $\rho$ is constant. However, here the effective density is constant, and consequently both density and pressure are functions of $r$. To find them, let us first write $\rho$ and $p$ in terms of the effective quantities. To do so we use \eqref{efro} and \eqref{efp} and ignore nonlinear contributions of $\eta c^2/\kappa$. The result is
\begin{equation}
\begin{split}
&\rho(r)\simeq \rho_{\text{eff}}+\frac{\eta c^2}{2 \kappa}\left(8{\rho_{\text{eff}}}\,\frac{p_{\text{eff}}}{c^2}+\,\rho_{\text{eff}}^2+3 \,\frac{p_{\text{eff}}^2}{c^4}\right)\\&
p(r)\simeq p_{\text{eff}}+\frac{\eta c^4 }{2\kappa}\Big(\rho_{\text{eff}}^2+3\,\frac{p_{\text{eff}}^2}{c^4}\Big)
\end{split}
\label{pres}
\end{equation}
substituting $\rho_{\text{eff}}=\rho_0$ and equation \eqref{ints}, into above equations one may easily find the exact form of $\rho(r)$ and $p(r)$ in terms of $r$. The result can be written as
\begin{equation}
\begin{split}
\rho(r)=&\rho_0+\frac{\eta c^2\rho_0^2}{2\kappa}\left(\frac{26\sqrt{(1-\alpha)(1-\alpha r^2/R^2})}{(\sqrt{1-\alpha r^2/R^2}-3\sqrt{1-\alpha})^2}\right)\\&+\frac{\eta c^2\rho_0^2}{2\kappa}\left(\frac{\alpha(21+5r^2/R^2)-26}{(\sqrt{1-\alpha r^2/R^2}-3\sqrt{1-\alpha})^2}\right)
\end{split}
\end{equation}
\begin{equation}
\begin{split}
&p(r)=\rho_0 c^2\left(\frac{\sqrt{1-\alpha r^2/R^2}-\sqrt{1-\alpha}}{3\sqrt{1-\alpha}-\sqrt{1-\alpha r^2/R^2}}\right)\\& +\frac{\rho_0^2 c^4 \eta}{2\kappa}\left(3\left(\frac{\sqrt{1-\alpha r^2/R^2}-\sqrt{1-\alpha}}{3\sqrt{1-\alpha}-\sqrt{1-\alpha r^2/R^2}}\right)^2+1\right)
\end{split}
\end{equation}
where, for the sake of simplicity, $\alpha$ is defined as 
\begin{equation}
\alpha=\frac{\kappa c^2 m_{\text{eff}} }{4\pi R}
\end{equation}
Note that as in GR, the central pressure obtained from (\ref{pres}) in the limit $r\rightarrow 0$, diverges when $\kappa c^2 m_{\text{eff}}/4\pi R\rightarrow 8/9$. Therefor the necessary condition for existence of this solution is $\kappa c^2 m_{\text{eff}}/8\pi R< 4/9$, which is reminiscent of the Buchdahl's theorem in GR. We found this result for a specific equation of state \eqref{eo1} in EMSG, while Buchdahl's theorem holds for all equation of states. 

\subsection{Relativistic pressureless stars: $p=0$}
It is interesting that in EMSG, pressureless stars can exists. We know that in GR, pressureless matter cannot be stable againt its own self-gravity. However, as it can be read from equation \eqref{efp}, {for a pressureless system the effective pressure is not necessarily zero.} Consequently the effective pressure may support the star against gravitational collapse. We set the pressure $p$ to zero in the main equations \eqref{two}, \eqref{one} and \eqref{three}. In this case from equation \eqref{three} one may straightforwardly find
\begin{equation}\label{new1}
b(r)=\left(1-\gamma \rho (r)\right)^{-2}
\end{equation}
where $\gamma$ is a constant of integration. We substitute this relation into \eqref{one} to obtain the following result for $f(r)$
\begin{equation}\label{new2}
f(r)=-\frac{\left(c^4 \,\eta  \,r^2 \rho (r)^2-2\right) \left(\kappa -c^2\, \eta \, \rho (r)\right)}{2 \left(2 c^2\, \eta  \,r \rho '(r)-c^2\, \eta \, \rho (r)+\kappa \right)}
\end{equation}
Finally we substitute \eqref{new2} into equation \eqref{two} to obtain a second order differential equation for $\rho(r)$. The result is 
\begin{eqnarray}\label{new3}
&\nonumber -2 c^6 \eta ^3 r \rho (r)^4-2 \eta  \Big(3 c^2 \eta  r \rho '(r)^2+\kappa  r \rho ''(r)+2 \kappa  \rho '(r)\Big)\\\nonumber&+c^4 \eta ^2 r \rho (r)^3 \Big(-c^2 \eta  r^2 \rho ''(r)+3 c^2 \eta  r \rho '(r)+5 \kappa \Big)\\\nonumber&+c^2 \eta  r \rho (r)^2 \Big(c^4 \eta ^2 r^2 \rho '(r)^2+\kappa  \left(c^2 \eta  r^2 \rho ''(r)-4 \kappa \right)\\&-6 c^2 \eta  \kappa  r \rho '(r)\Big)+\rho (r) \Big(2 c^4 \eta ^2 \kappa  r^3 \rho '(r)^2\\&+c^2 \eta  (4 \eta +3 \kappa ^2 r^2) \rho '(r)+r(2 c^2 \eta ^2 \rho ''(r)+\kappa ^3)\Big)=0\nonumber
\end{eqnarray}
this non-linear equation can be solved numerically, and we could not find any analytic solution for it. In order to find an exact solution, let us restrict ourselves to a constant density case, i.e. $\rho'(r)=0$. In this case equation \eqref{new3} can be written as
\begin{equation}
\Big(1-\frac{2\eta c^2}{\kappa}\rho\Big)\Big(1-\frac{\eta c^2}{\kappa}\rho\Big)=0
\end{equation}
One of the solutions is larger than the maximum density allowed in the early universe. Therefore we choose the second solution given by 
\begin{equation}\label{new4}
\rho(r)=\frac{\kappa}{2\eta c^2}
\end{equation}
substituting \eqref{new4} into equation \eqref{new1} we arrive at $b(r)=\gamma^*$, where $\gamma^*$ is a constant which should be determined using the boundary conditions. We know that at the surface of the star, the metric is given by the Schwarzschild metric. Consequently we have
\begin{equation}\label{new5}
b(r)=1-\frac{\kappa c^2 M^{*}}{4\pi R^{*}}
\end{equation}
 where $M^{*}$ and $R^{*}$ are the mass and radius of the star respectively. In fact by matching the interior and exterior solutions it is easy to show that $M^{*}=m_{\text{eff}}(R^{*})=\frac{\pi R^{*3} \kappa}{2\eta c^2}$. Finally equation \eqref{new2} yields
\begin{equation}
f(r)=1-\frac{\kappa^2}{8\eta}r^2
\end{equation} 
 
 It is important to mention that the density of this star is exactly equal to the maximum energy density which appear in the early universe \cite{us1}. Therefore this solution presents a very dense star. Naturally, if the radius of this star is smaller than the Schwarzschild radius, then it can be considered as a black hole. We stress again that the metric outside the star is given by the Schwarzschild space-time. This happens since there is no difference between EMSG and GR in the vacuum. Interestingly, this behavior directly implies that the Birkhoff's theorem holds also in EMSG. In this case, we have found the exact solution for the metric components and the matter distribution inside a pressureless black hole with constant density in EMSG.

In fact as has been shown in \cite{us1}, this solution is reminiscent of Einstein's cosmological model where the cosmological constant $\Lambda$ introduced to support a static cosmological model. {More specifically, quite similar to the above solution, existence of $\Lambda$ appears as a negative effective pressure and prevents the gravitational collapse.} However we know that Einstein model is not stable. Let us check the stability of our solution against small perturbations. To shed light on the stability of this solution, and for the sake of simplicity, we consider the corresponding star in the weak field limit. In this case, using the mathematical similarity between GR and EMSG, we can say that the mathematical form of the Newtonian governing equations, i.e. the continuity, Euler and Poisson equations, do not change in EMSG. The only difference is to replace $\rho$ and $p$ in the standard case with the corresponding effective quantities. Therefore our main equations to study the stability of the system are
\begin{equation}
\frac{\partial \rho_\text{eff}}{\partial t}+\nabla \cdot\left(\rho_{\text{eff}} \mathbf{u}\right)=0
\label{cont}
\end{equation}
\begin{equation}
\frac{d\mathbf{u}}{dt}+\frac{\nabla p_\text{eff}}{ \rho_\text{eff}} + \nabla \Phi=0
\label{euler}
\end{equation}
\begin{equation}
\nabla ^2 \Phi =\frac{1}{2}\kappa c^4 \rho _\text{eff}
\label{poisson}
\end{equation}
Where $\mathbf{u}$ is the velocity field of the fluid, and is zero for the background solution. {In the following we study the stability of the solution using the both Eulerian and Lagrangian formulations.}

Using the Eulerian description of the small perturbations, we perturb the physical quantities collectively shown as $Q\rightarrow Q+Q_1$. Where subscripts "$1$" stands for first order perturbations. Substituting these perturbations within equations \eqref{cont}-\eqref{poisson}, and keeping only the first order terms, we arrive at 
\begin{equation}
\frac{\partial \rho_\text{eff1}}{\partial t}+\nabla \cdot\left(\rho_{\text{eff}} \mathbf{u_1}\right)=0
\label{cont1}
\end{equation}
\begin{equation}
\frac{d\mathbf{u_1}}{dt}+c_s^2\frac{\nabla p_\text{eff1}}{ \rho_\text{eff}} + \nabla \Phi_1=0
\label{euler1}
\end{equation}
\begin{equation}
\nabla ^2 \Phi_1 =\frac{1}{2} \kappa c^4 \rho _\text{eff1}
\label{poisson1}
\end{equation}
where $p_{\text{eff1}}=p_1/2$, $\rho_{\text{eff1}}=\rho_1/2$, and the effective sound speed $c_s$ is defined as 
\begin{equation}
c_s^2=\frac{dp_{\text{eff}}}{d\rho _{\text{eff}}}
\end{equation}
{Of course this effective speed parameter does not belong to any physical transform of matter or energy.} It should be noted that the sound speed for the background system, which is pressureless, is zero. It is easy to verify that $c_s^2=-c^2$. By differentiating equation \eqref{cont1} with respect to time, and using equations \eqref{euler1} and \eqref{poisson1}, we find
\begin{equation}
\frac{\partial^2 \rho_1}{\partial t^2}-c_s^2 \nabla^2 \rho_1-\frac{1}{2}\kappa c^4  \rho_{\text{eff}}\,\rho_1=0
\label{pert2}
\end{equation}
Now as usual let us assume a Fourier mode for the perturbation $\rho_1$ as follows
\begin{equation}
\rho_1(\mathbf{r},t) =\rho_a e^{i \left(\mathbf{k}\cdot\mathbf{r}-\omega t\right)}
\end{equation}
by substituting this solution within \eqref{pert2}, we find the following dispersion relation {for the propagation of the plane waves}
\begin{equation}
\omega ^2 = c_s^2 k^2 - \frac{1}{2}\kappa c^4 \rho_{\text{eff}}
\end{equation}
It is clear that since $c_s^2<0$ and $\rho_{\text{eff}}>0$ the right hand side is always negative. This means that all wavelengths are unstable. Consequently, this stellar solution is unstable and is not interesting from observational point of view. {However, it is interesting that, not only this solution is reminiscent of Einstein's cosmological model, also is unstable as well.}

Although the above stability analysis is enough to rule out this solution, it is also instructive to use the Lagrangian description for the evolution of the radial adiabatic perturbations. In fact, the radial stability analysis of the star can be done analytically for this case. In the following we briefly prove that the star is also unstable against radial perturbations. Assuming that the Lagrangian displacement is $\xi$ and the velocity is $\mathbf{u}=(u(r,t),0,0)$, the perturbed Euler equation is written as
\begin{equation}
\Delta \left( \frac{du}{dt}+\frac{p_\text{eff}'}{ \rho_\text{eff}} + \Phi'\right)=0
\end{equation}
where the Lagrangian perturbation operator $\Delta $ and the Eulerian perturbation operator $\delta$ are related as 
\begin{equation}
\Delta=\delta+ \mathcal{\xi}\cdot\nabla
\end{equation}
for more details we refer the reader to \cite{shapiro}. Again we benefit the mathematical similarity between EMSG and GR. In this case it is easy to verify that the eigenvalue equation for the radial perturbations in a spherical star is given by (see equation (6.5.6) in \cite{shapiro})
\begin{equation}
\frac{d}{dr}\Big(\rho_{\text{eff}}\,\frac{c_s^2 }{r^2}\frac{d}{dr}(r^2 \xi)\Big)-\frac{4 }{r^2}\frac{d p_{\text{eff}}}{dr}\xi+\omega^2\rho_{\text{eff}}\,\xi=0
\end{equation}
where $\xi(r,t)$ is the radial component of vector $\mathbf{\xi}$, and we have assumed a time dependency $e^{i\omega t}$ for it. Keeping in mind that $p_{\text{eff}}$ is constant, the above equation can be rewritten as
\begin{equation}
\frac{d}{dr}\Big(\frac{1}{r^2}\frac{d}{dr}(r^2 \xi)\Big)+\frac{\omega^2}{c_s^2}\xi=0
\end{equation}
{Fortunately, this differential equation can be simply solved to arrive at}
\begin{equation}
\xi(r,t)=\Big(\alpha\, j_1\Big(\frac{r \omega }{c_s}\Big)+\beta \,y_1\Big(\frac{r \omega }{c_s}\Big)\Big)e^{i \omega t}
\end{equation}
where $\alpha$ and $\beta$ are constants of integration, and $j_1$ and $y_1$ are the spherical Bessel functions of the first and second kinds respectively. On the other hand since, $c_s^2<0$, we can not find any stable mode ($\omega^2>0$), and all frequencies are unstable. In fact, for the same reason as in the Eulerian description, i.e. $c_s^2<0$, all radial perturbations are unstable in the Lagrangian description.

\section{Numerical solutions}
\label{numeric}
\begin{figure}
\includegraphics[scale=0.8]{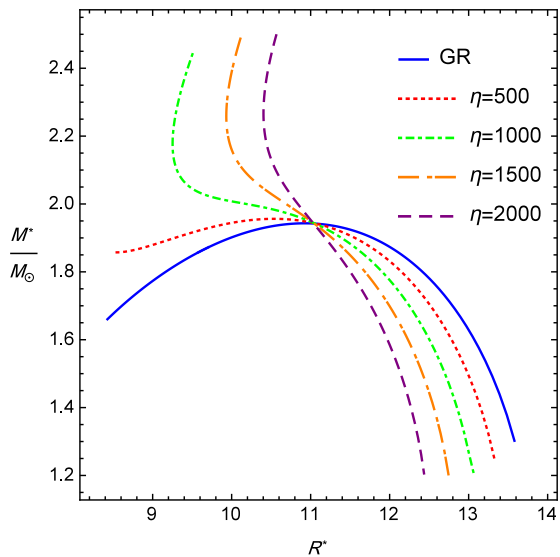}\caption{The mass-radius relation for different values of $\eta$. {The blue solid curve belongs to GR}. Curves terminate when there is no physical solution for the governing equations. {The mass has been scaled in terms of solar mass $M_{\odot}$}.}
\label{fig1}
\end{figure}
In this section we solve the field equations \eqref{two}, \eqref{one} and \eqref{three} for more realistic stars. To do so it is necessary to choose an appropriate set of boundary conditions. At the center, it is clear from \eqref{two} and \eqref{one} that $f(0)=1$. On the other hand, we solve the equations for a wide range of central pressure, $p(0)=p_c$. Therefore we know the central pressure and density, i.e. $\rho(0)=\rho_c$. It is necessary to mention that with a time reparametrization we can set $b(0)=1$. In fact, it turns out that density profile and the final mass and the radius of the star remain unchanged by changing $b(0)$.  

Therefore we know all the three initial conditions necessary to solve the equations. On the other hand in order to find the mass and radius of the star, the field equations \eqref{two}, \eqref{one} and \eqref{three} are integrated numerically outward up to the star radius $R^*$, where the pressure gets zero, i.e. $p(R^*)=0$. On the other hand, our numerical solution should match the external solution for $r>R^*$. By matching the solution we can find the mass of the star, i.e. $M^*$. Fortunately, the external solution is exact and given by Schwarzschild metric, for which $b(r)=f(r)=1-\kappa c^2 M^*/4\pi r$. {We emphasize again that the mass appeared in the external metric, is obtained from $\rho_{\text{eff}}$ and not $\rho$.} Imposing the Darmois-Israel matching condition for a spherically symmetric spacetime, the matching conditions can be written as
\begin{equation}\label{new10}
[b(r)]=[f(r)]=0
\end{equation}
where $[...]$ is the jump across the surface $r=R^*$ defined as 
$[T]=T^{+}(R^{*})-T^{-}(R^{*})$ where $+$ and $-$ means the quantity $T$ evaluated outside and inside the surface respectively. On the other hand, using equations \eqref{two} and \eqref{one}, one can show that the radial derivative of the metric components are not continuous across the surface. More specifically we have
\begin{equation}
[f']=\kappa c^2 R \rho_{\text{eff}}^{-}(R^{*}), ~~~~~[b']=-\kappa R^2 p_{\text{eff}}^{-}(R^{*})
\label{sur}
\end{equation}
Albeit in order to find the mass of the star, i.e. $M$, we only need to use \eqref{new10}. Notice that for the same situation in GR, we have $[b']=0$. Therefore discontinuity in $[b']$ seems as a new feature in EMSG. However, as we already mentioned the pressure vanishes at the surface of the star. Therefore using the definition of $p_{\text{eff}}$ and equation \eqref{sur}, we have 
\begin{equation}
[b']=\frac{\eta c^4}{2}R^2\rho^{-}(R^*)^2
\end{equation}
On the other hand, for ordinary matter distributions which obey the standard polytropic EOS, the matter density vanishes everywhere the pressure is zero, for example see \eqref{eoso}. Therefore for these systems as in GR we have $[b']=0$. In other words, for well-known models for neutron stars we have $[b']=0$.

On the other hand for exotic quark stars descried in the subsequent sections, the matter density can be nonzero at $r=R^*$ while the pressure is zero. In this case we have a discontinuity in $[b']$ in EMSG. However it is well-known that the physics and the conditions for the existence of these stars, are still hypothetical and unproven from observational point of view. Consequently it is unlikely to expect a direct observational evidence for this new feature in EMSG. Of course, a careful and detailed analysis of junction conditions in EMSG, would help to clarify this issue. We leave this point as a subject of study for future works.

It is necessary to mention that in order to solve the equations we need the EOS of the star. In the following we study two different EOS, which are widely used in the literature to investigate the internal structure of relativistic stars.

\begin{figure}
\includegraphics[scale=0.8]{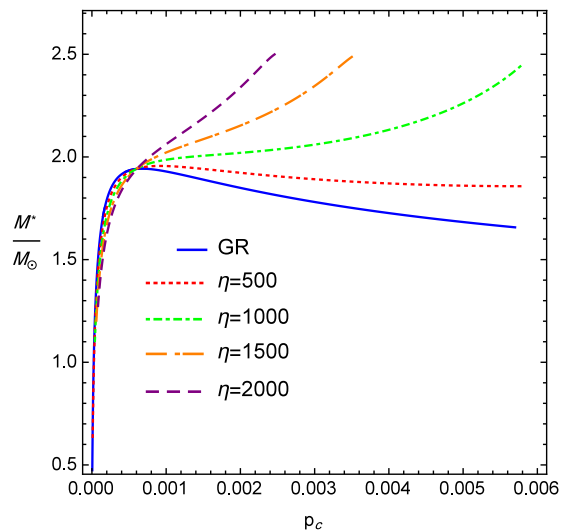}\caption{ The mass of the star with respect to the central pressure $p_c$ ($\text{km}^{-2}$) for the same values of $\eta$ presented in the figure \ref{fig1}.} \label{fig2}
\end{figure}
\subsection{Polytropic stars}
Hereafter for simplicity in numerical integration we work in units where $c=G=1$. It is well known that the EOS of neutron star matter is still a challenge topic at present. However, for describing matter inside a neutron star, {at least at the first approximation}, it is convenient to follow Damour and Esposito-Farese work \cite{damour} in which a polytropic EOS has been introduced. This EOS is given by 
\begin{equation}
\rho=\dfrac{p}{\Gamma-1}+\left(\dfrac{p}{K\rho_0^{1-\Gamma}}\right)^{1/\Gamma}
\label{eoso}
\end{equation}
where the exponent $\Gamma$ and polytropic constant $K$ are dimensionless parameters. We use two cases for these parameters introduced in \cite{damour} as $K=0.0195,\,\Gamma=2.34$ and $K=0.00936,\,\Gamma=2.46$. However the main results are qualitatively the same. Therefore we only report the former case when $\rho_0=1.66\times 10^{17} \text{kg}\,\text{m}^{-3}$. In order to integrate the governing equations numerically, using the procedure introduced in the previous section, we choose the central pressure $p_c$ in the interval $6\times10^{29}-1.2\times10^{33}\, \text{Pa}$. This choice yields to the central energy in the interval $2.3\times10^{15}-1.59\times10^{17}\, \text{kg}\, \text{m}^{-3}$ . These values are consistent with the typical values attributed to the central pressure and density of neutron stars.

The results have been shown in Fig. \ref{fig1} and \ref{fig2}. Fig. \ref{fig1} shows the mass-radius relation for different values of $\eta$. The blue solid curve belongs to GR. This curve coincides with the results presented in \cite{gama2.34}. As expected by increasing $\eta$ the deviation between GR and EMSG increases. It should be noted that the central pressure $p_c$ increases when moving from right to left on each curve in this figure. 

On the other hand, Fig \ref{fig2} illustrates the mass of the star with respect to $p_c$. As expected at larger central pressures the deviation between two theories gets more evident. It is necessary to mention that there are two restrictions which must be satisfied, and curves terminate when these conditions are violated. The first condition is  that the sound speed at the center must be smaller than the speed of light. On the other hand, the second derivative of the pressure should be negative in the center of the compact star. As we already mentioned, we have an effective pressure which plays role instead of pressure itself. Consequently we need $d^2 p_{\text{eff}}/dr^2<0$ at $r=0$. As we shall see, this condition restricts the magnitude of the central pressure. To see this fact more clearly let us expand the metric components and effective quantities around the center as follows
\begin{equation}
\begin{split}
&f(1)\simeq 1+f_2 r^2~~~~,~~~~b(r)\simeq 1+b_2 r^2\\&
\rho_{\text{eff}}(r)\simeq \rho_e+\rho_2 r^2,~~~~p_{\text{eff}}(r)\simeq p_e+p_2 r^2
\end{split}
\end{equation}
now substituting these expansions into the main equations \eqref{two}, \eqref{one} and \eqref{three}, we find three equations for three coefficients, i.e. $f_2$, $b_2$ and $p_e$. Notice that $\rho_e$ does not appear in these equations. Finally we find the coefficients in terms of $\rho_e$ and $p_e$ as follows
\begin{equation}
\begin{split}
&f_2=\frac{1}{3}\kappa \rho_e\\&
b_2=\frac{\kappa}{6}(\rho_e+3 p_e)\\&
p_2= -\frac{\kappa}{12}(3 p_e^2+4 \rho_e p_e+\rho_e^2)
\end{split}
\end{equation}
therefor the above mentioned condition for the existence of compact stars, i.e. $p_2<0$, can be written as
\begin{equation}\label{cnew}
3 p_e^2+4 \rho_e p_e+\rho_e^2>0
\end{equation}
where the central values of the effective quantities are related to the corresponding values for the pressure and density of the system as follows
\begin{equation}
\begin{split}
&\rho_e=\rho_c-\frac{\eta}{2 \kappa}(8{\rho_c}\,p_c+\,\rho_c^2+3 \,p_c^2)\\&
p_e=p_c-\frac{\eta }{2\kappa}(\rho_c^2+3\,p_c^2)
\end{split}
\end{equation}
Now assuming that the central density is smaller than the allowed maximum density in the theory, i.e. $\rho_{\text{max}}=\kappa/2\eta$, one can easily show that the condition \eqref{cnew} is satisfied if

\begin{equation}
p_c< \dfrac{\kappa}{3\eta}-\frac{\rho_c}{3}
\end{equation}
In fact, our numerical solutions show that this condition also guarantees that $dp_{\text{eff}}/dr<0$ everywhere throughout the star. We have taken into account both of these restrictions in the solutions.
\begin{figure}
\includegraphics[scale=0.8]{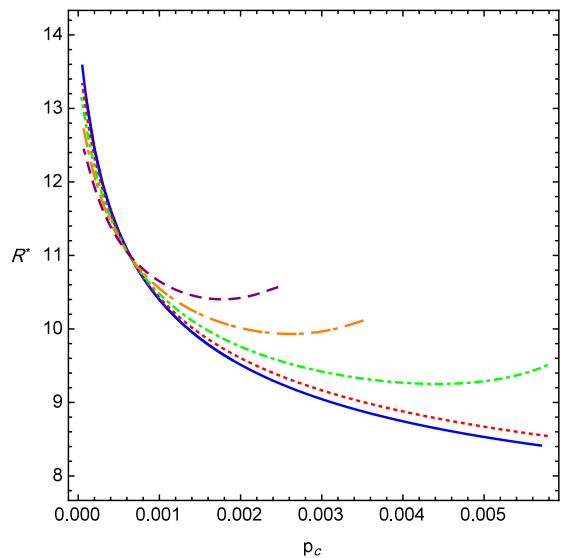}\caption{The radius of the star $R^* (\text{km})$ with respect to the central pressure $p_c$ ($\text{km}^{-2}$). Different colors belong to the different values of $\eta$ presented in Fig. \ref{fig1}. The blue solid curve belongs to GR.}
\label{fig6}
\end{figure}
Interestingly, as it is clear form Fig. \ref{fig2} for stars with the same central pressures in the interval $p_c \lesssim 5.8 \times 10^{-4}\text{km}^{-2}$, EMSG leads to smaller masses for neutron stars compared with GR. On the other hand, in higher central pressures, EMSG leads to larger masses. In Fig. \ref{fig6} we have shown the radius of the star with respect to the central pressure. It is clear that for $p_c< 6\times 10^{-4}\text{km}^{-2}$ ($p_c> 6\times 10^{-4}\text{km}^{-2}$) the radius of the star is smaller (larger) than in GR. This is interesting in the sense that, within GR, this EOS leads to maximum mass $M^*\simeq 1.9 M_{\odot}$, which is smaller than the observed value $2.01\pm 0.04 M_{\odot}$ \cite{ns}. In other words, EMSG can recover the observed value without invoking stiffer EOS.

It is also interesting to mention that, as it is clear Fig. \ref{fig2}, for each $\eta$ there is a specific $p_c$ for which the mass of the star is equivalent to a star in GR with different radius and internal properties. On the other hand, as seen in Fig. \ref{fig1}, for each value of $\eta$, always exists a star in EMSG with a same mass and radius in GR. However the central pressure of these stars and their density profiles are not the same. Although in both figures it seems that all curves intersect the GR curve in a same point, it turns out that this is not the case.

\begin{figure}
\includegraphics[scale=0.8]{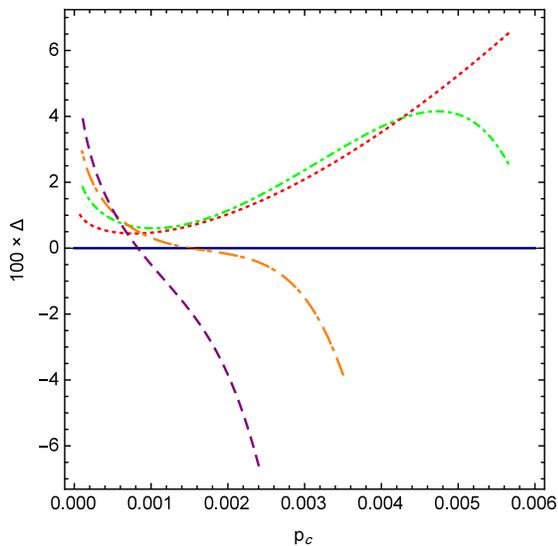}\caption{Relative difference between density parameter $C$ in EMSG and GR with respect to $pc$ ($\text{km}^{-2}$). Different colors belong to different values of $\eta$ used in previous figures.}
\label{fig3}
\end{figure}

It should be noted that for low $p_c$, where EMSG gives smaller masses compared with GR, the radius of the stars is also smaller than in GR. In this case it can be shown that stars are more compact in EMSG. To do so let us define parameter $C$ as a measure of the mean density as: $C=M^{*}/{R^{*}}^3$. We have plotted the relative difference between $C$ in EMSG and GR with respect to central pressure, i.e $\Delta=(C_{EMSG}-C_{GR})/C_{GR}$ in Fig. \ref{fig3}. It is clear form this figure that for small $p_c$, and independent from the magnitude of $\eta$, the mean density defined by $C$, in EMSG is larger than GR. Existence of more compact stars in EMSG is not surprising in the sense that the main feature of the theory is to avoid singularities. However it is clear that when the central pressure and the free parameter $\eta$ are relatively large, stars can be less compact in EMSG. 
\begin{figure*}[!]
\includegraphics[scale=0.7]{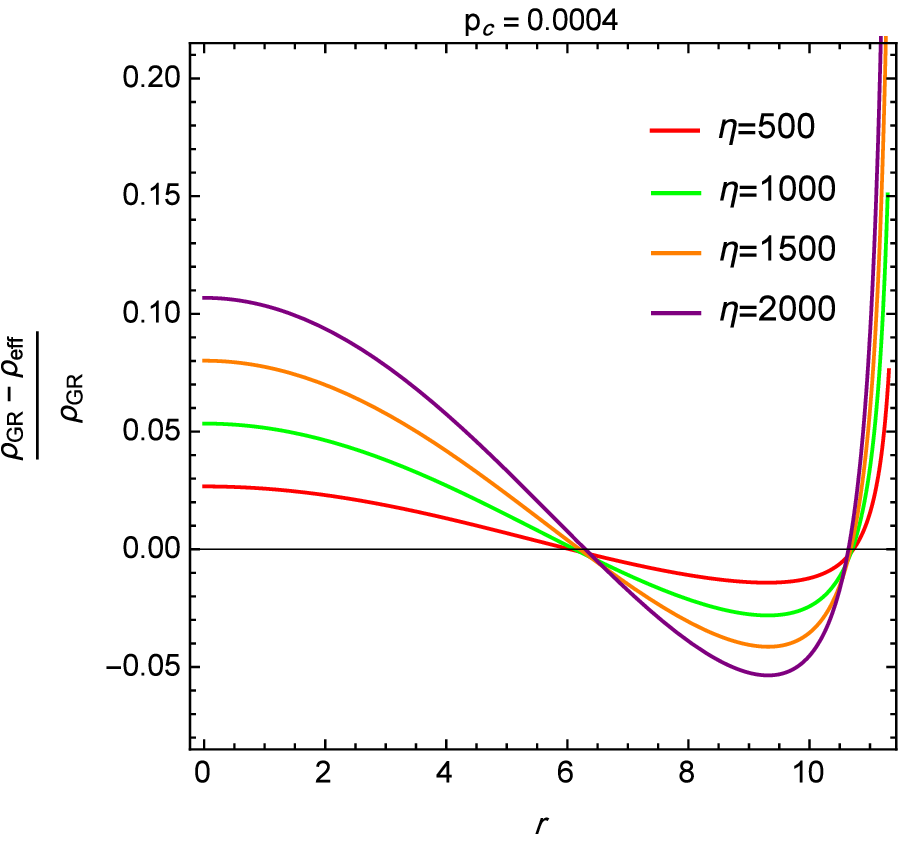}
\hspace*{10mm}
\includegraphics[scale=0.7]{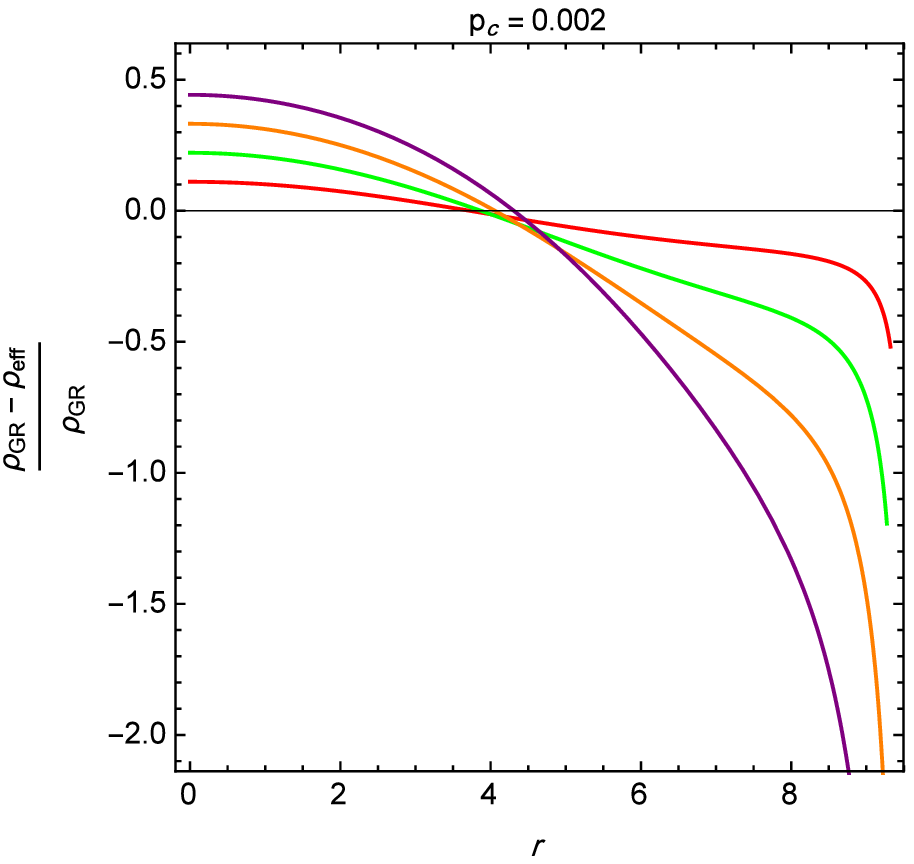}
\vspace*{10mm}
\includegraphics[scale=0.7]{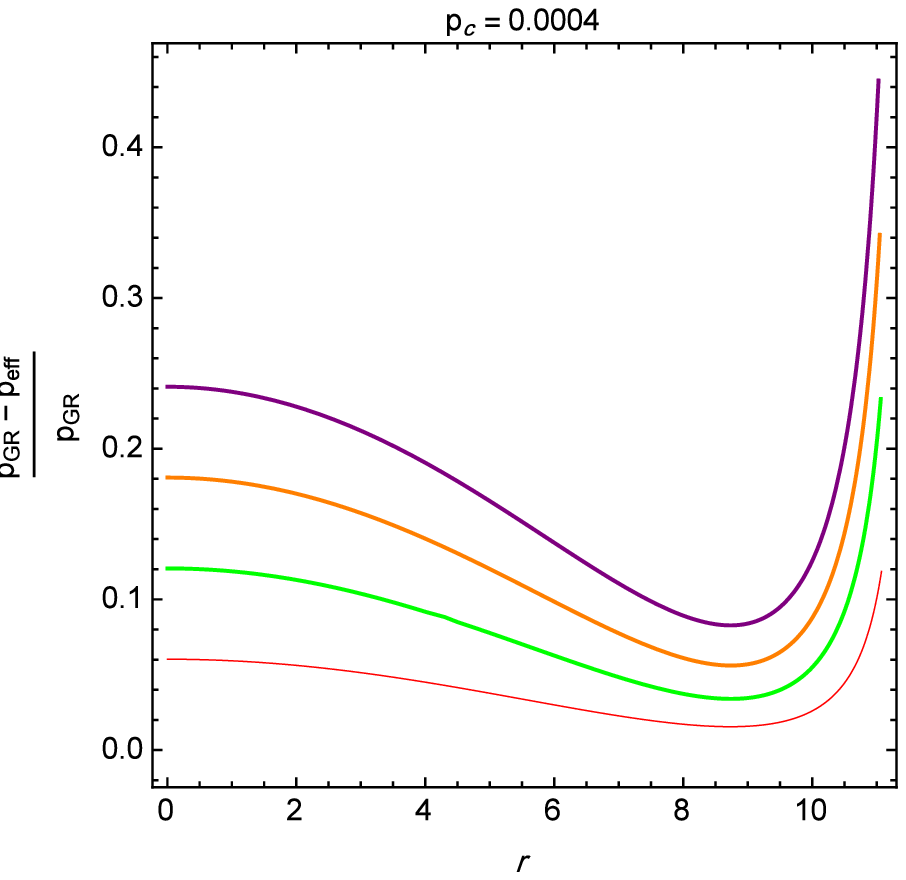}
\hspace*{10mm}
\includegraphics[scale=0.7]{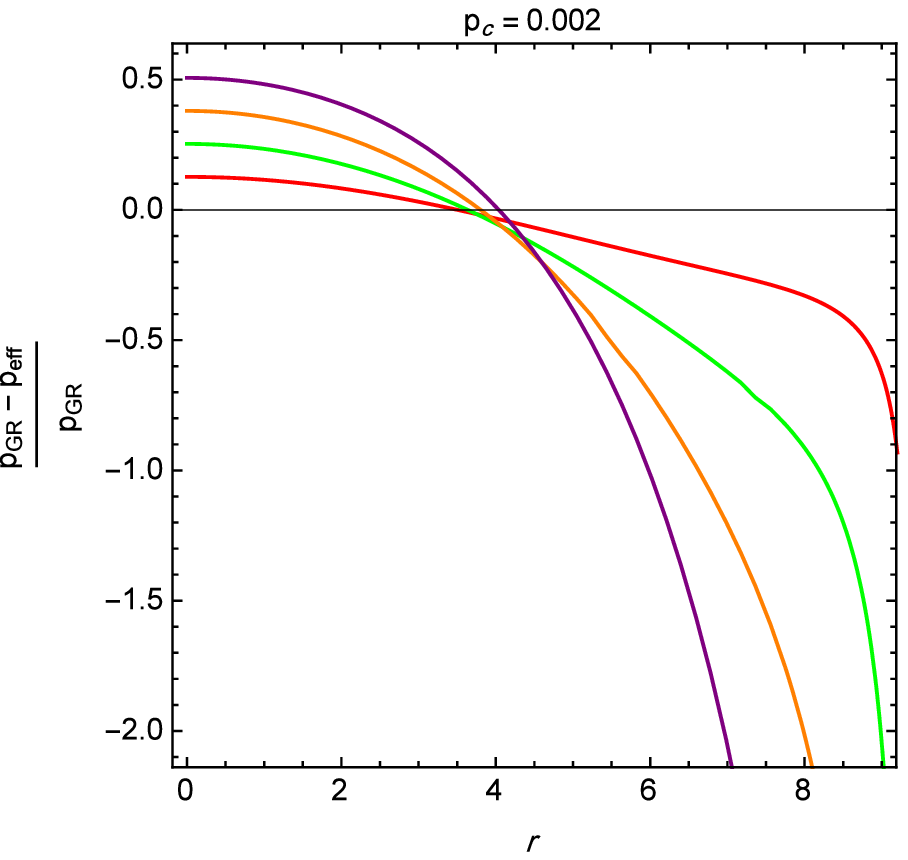}
\vspace*{-10mm}
\caption{\textit{Top panels:} Relative difference between $\rho_{\text{GR}}$ and $\rho_{\text{eff}}$ for two different values of $p_c$, i.e. $p_c=0.0004$ and $0.002$, with respect to distance from center of the star. \textit{Bottom panels:} The corresponding relative difference between $p_{\text{GR}}$ and $p_{\text{eff}}$ with respect to $r$.
}
\label{fig11}
\end{figure*}

It is important mentioning that for two stars with the same central pressure, one in GR and one in EMSG, the density and pressure are not necessarily smaller in EMSG. More specifically, depending on the radius $\rho_{\text{eff}}$ ($p_{\text{eff}}$) can be smaller or even larger than the corresponding values in GR. In the top panels of Fig. \ref{fig11} we have illustrated relative difference between density in GR, i.e. $\rho_{\text{GR}}$, and the effective density in EMSG for two different values of central pressure with respect to distance from the center of the star. It is clear that at larger distances the effective density gets larger than $\rho_{\text{GR}}$. Also when $p_c$ is larger, the interval in which $\rho_{\text{eff}>}\rho_{\text{GR}}$ gets wider. On the other hand, as expected, larger $\eta$ leads to larger differences between two theories.

In the bottom panels of Fig \ref{fig11}, we have shown the relative difference between pressure of the star in GR, i.e. $p_{\text{GR}}$, and the effective pressure in EMSG, for two different values of $p_c$. It is clear again that for large $p_c$, at larger radii the effective pressure is larger than $p_{\text{GR}}$. On the other hand when $p_c$ is small, the effective pressure remains smaller than $p_{\text{GR}}$ for all chosen values of $\eta$. 

The overall outcome of these differences is that the star, depending on the values of $\eta$ and central pressure $p_c$, the mean density defined by $C$ can be smaller or larger than in GR.

As our final remark in this section, it should be noted that, as expected, the behavior of EMSG inside compact stars depends on the EOS and internal properties. In the subsequent section, as another example, we briefly show that EMSG increases the mass of quark stars for all values of $\eta$ and $p_c$. Furthermore these stars are less compact in EMSG compared with GR.
\begin{figure*}[!]
\includegraphics[scale=0.7]{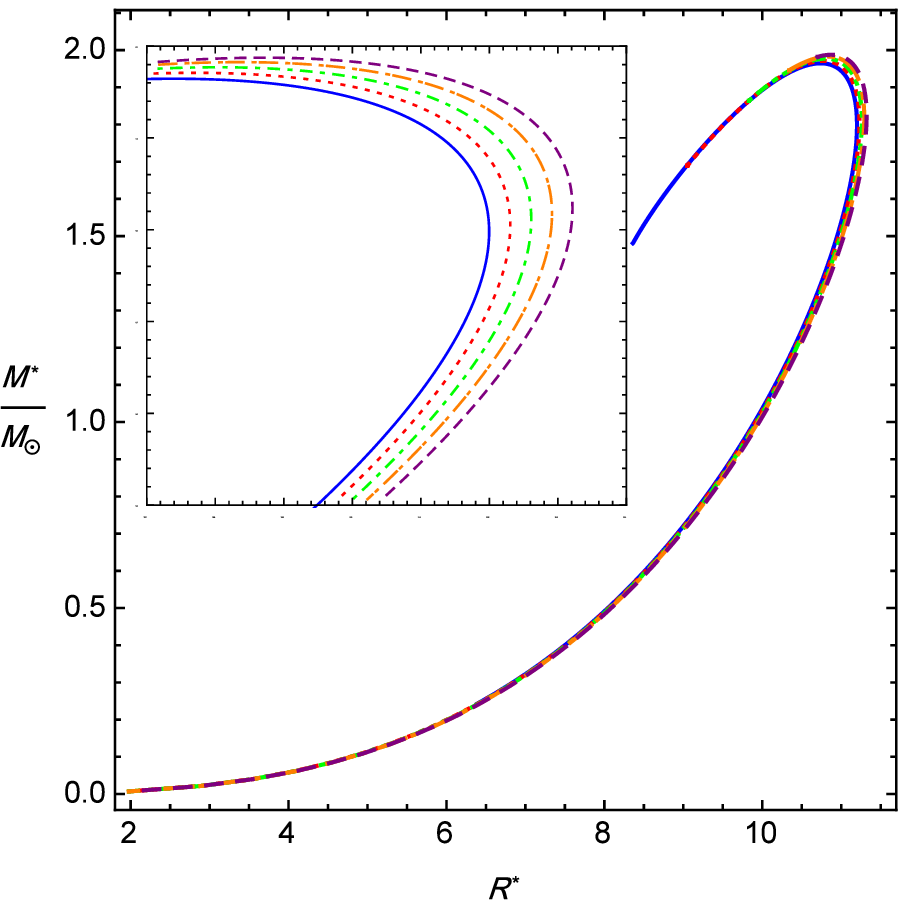}
\hspace*{10mm}
\includegraphics[scale=0.7]{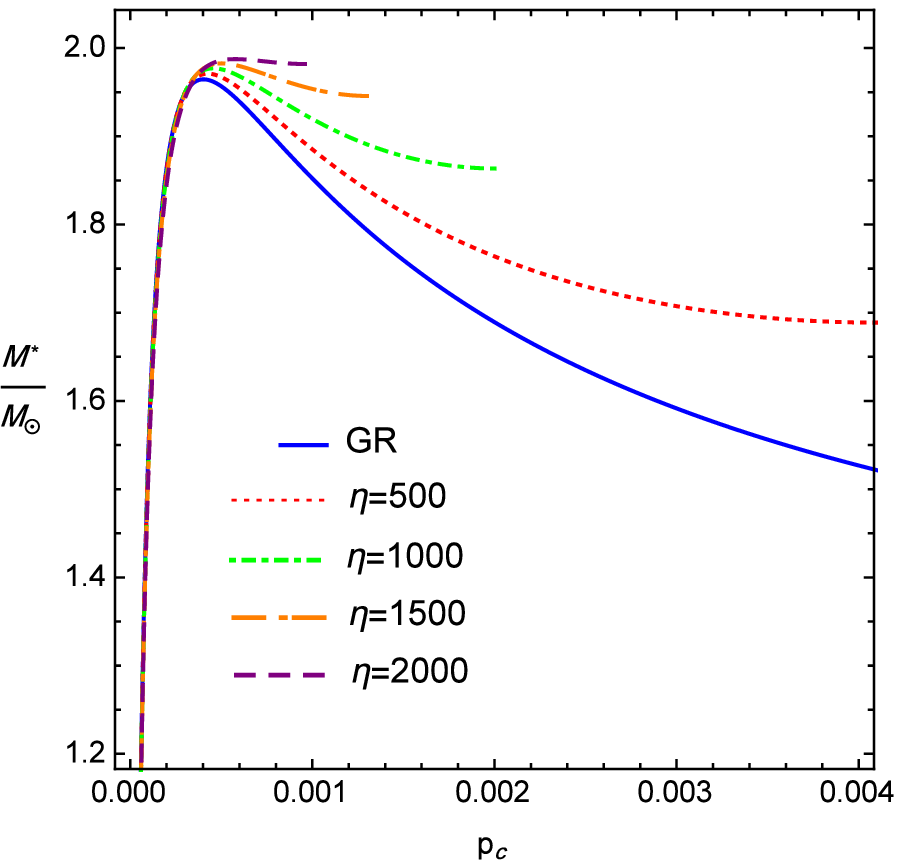}
\hspace*{0mm}
\includegraphics[scale=0.7]{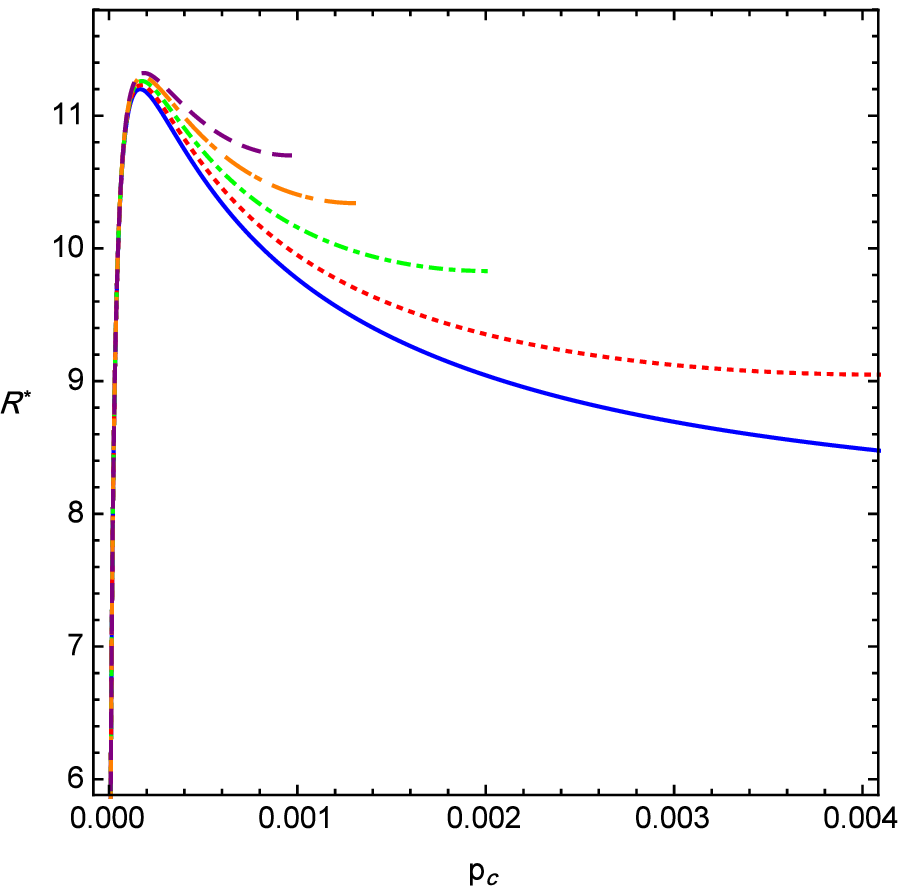}
\hspace*{10mm}
\includegraphics[scale=0.7]{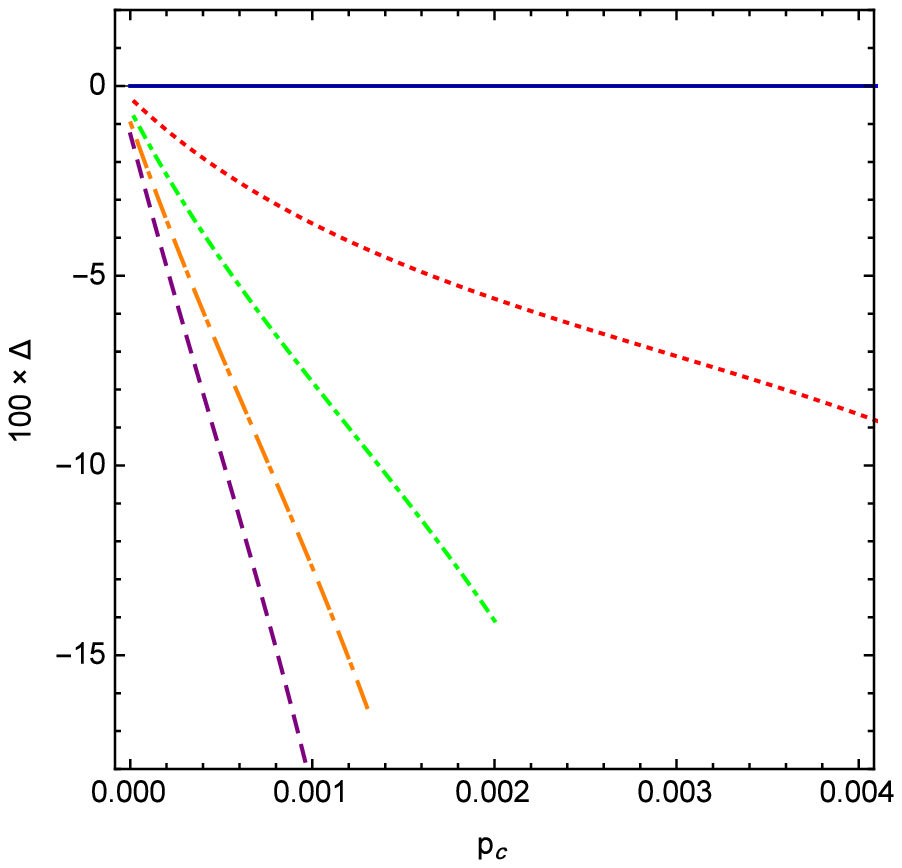}
\caption{In the all panels the blue curve belongs to GR and other colors demonstrates different values for $\eta$ given in the top right panel. \textit{Top panels:} the left panel shows the mass-radius relation of strange quark stars in EMSG, and the right panel shows the radius of the star with respect to $p_c$. \textit{Bottom panels:} The left panel represents the radius of the star with respect to $p_c$, and the right panel shows the relative difference between the density parameter $C$ of quark stars in GR and EMSG.
}
\label{fig12}
\end{figure*}

\subsection{Strange quark stars:}
By assuming that the strange quark matter consisting of massless quarks is the ultimate ground state of matter, then compression of matter to high temperature and density may convert the entire star into strange quark matter. From observational point of view it is very difficult to distinguish between these stars and normal neutron stars, for more details about these stars we refer the reader to \cite{quark2}. In the first order perturbation theory in quantum chromodynamics (QCD), the EOS of quark stars is given by \cite{quark1}
\begin{equation}
\rho=3\,p+4B
\label{quarkEQS}
\end{equation}
where B (the bag constant) is the difference between the energy density of the perturbative and non-perturbative QCD vacuums. At zero pressure and temperature $B$ holds quark matter together and guarantee the existence of the strange quark stars. Here we adopt the typical value of the bag constant given by $B\simeq 9.6\times10^{17}\text{kg}\text{m}^{-3}$ \cite{quark3}. With this choice, at the surface of star where $p=0$, we have $\rho=4\,B$. In other words, assuming that the density $\rho$ is a decreasing function, $\rho$ must be larger than $4\,B$ everywhere in the star.

The numerical results for this case, has been summarized in Fig. \ref{fig12}. In this figure blue curves belong to GR, and other colors belong to different values of $\eta$ in EMSG, and are equal to values which have been used for polytrops. The top left panel shows the mass-radius relation. For these stars EMSG leads to larger masses for all values of central pressure and $\eta$. However the difference is not too significant. We have zoomed to the curves and plotted it as a box inside the top left panel. For this EOS, for the same reason we already mentioned for polytopic stars, curves terminate when the following condition fails
\begin{equation}
p_c<\frac{\kappa}{6\eta}-\frac{2\,B}{3}
\end{equation}
 In the top right panel, the mass of the star is plotted with respect to the central pressure. It is clear that by increasing $\eta$, EMSG leads to larger masses. Furthermore it is seen in the bottom left panel that the radius of stars is also larger in EMSG. However, the mean density parameter $C$ of quark stars are significantly smaller in EMSG, see the bottom right panel in Fig. \ref{fig12}. Comparing this panel with Fig. \ref{fig3} it turns out that EMSG is more efficient in quark stars to reduce the mean density defined by $C$. Another significant difference is that EMSG prevents quark stars with high central pressures. In other words, schematically, we see that for larger $\eta$ all curves terminate in a relatively small $p_c$ compared with allowed pressures in stars in GR.

Before closing this section, it is worth mentioning that the EOS (\ref{quarkEQS}) has an interesting behavior in EMSG. In fact let us assume that the EOS is linear and given by $\rho= w \,p$ where $w$ is the equation of state parameter. In this case it is easy to show that in EMSG we have, effectively, a same EOS, i.e $\rho_{\text{eff}}=w\,p_{\text{eff}}$, provided that $w=3$ or $w=-1$. Therefore for a radiation dominated star with $w=3$, there would be no difference between mass-radius relation in GR and EMSG. However in strange quark stars although $w=3$, there is also a constant, i.e. $B$ in the EOS. In this case the EOS does not keep its shape, i.e. equation (\ref{quarkEQS}), and can be expressed as
\begin{equation}
\rho_{{eff}} = 3 p_{\text{eff}} + 4 B +\frac{8 \eta B}{\kappa} \left(2 B + p \right)
\label{b2}
\end{equation}
Consequently it is natural to expect small deviations from mass-radius relation in GR. {In this case, as it is clear form \eqref{b2}, the bag constant $B$, measures the magnitude of the deviations between EMSG and GR.}
\section{Conclusions}
In this paper two kind of compact stars have been investigated in the context of Energy-momentum-squared gravity. This theory leads to a nonsingular cosmology in the early universe, and its corrections to the standard gravity appear only in the high energy regime. The cosmological consequences of the theory has been investigated in \cite{us1}. The main aim in the current paper is to study the possible consequences of EMSG in compact stars, where we expect deviations from GR. By defining appropriate effective quantities, EMSG's field equations get similar to those of GR. This similarity induces substantial simplicity in calculations and interpretations. 

We found two exact stellar solutions. The first solution corresponds to a star with constant $\rho_{\text{eff}}$. In this case we could exactly find the radial dependency of $\rho(r)$ and $p(r)$. 

 In our second analytic solution, we investigated a pressureless star in the context of EMSG. As we know, such a star can not exist in GR. Although pressureless stars can exist in EMSG, we show that these relativistic stars are not stable. We checked the stability of these stars using both Lagrangian and Eulerian descriptions, and both methods confirmed that pressureless stars are unstable to local and radial perturbations. From this perspective, this stellar solution is reminiscent of Einstein's cosmological model.

As an attempt to include more realistic stars in this study, we found two numeric solutions. More specifically, we solved the field equations for two different EOSs: the polytopic and strange quark stars. In the case of polytropic stars it turned out that neutron stars with masses larger than $2\, M_{\odot}$ are allowed in EMSG. Therefore EMSG can explain recent high mass neutron star observations without invoking exotic EOSs. It is interesting that for small central pressures, EMSG leads to smaller masses compared to corresponding stars in GR. Furthermore, in some cases, depending on the internal structure of the stars, EMSG leads to more compact stars compared with GR, and in some situations lead to less compact stars.

On the other hand in the case of strange quark stars we found that masses are always larger in EMSG, and the stars are less compact. Also we showed that large central pressures, compared with the corresponding quark stars in GR, are not allowed in EMSG.

For future studies it would be interesting to study the slowly rotating stars in the context of EMSG. Also the Palatini formulation of EMSG may lead to interesting consequences in the stellar and cosmological issues.

\section{Acknowledgments}
We would like to appreciate the anonymous referee for very helpful comments. Also we wish to thank Fatimah Shojai for constructive discussions and comments. MR thanks Shahab Shahidi for reading this paper and mentioning important comments. This work is supported by Ferdowsi University of Mashhad under Grant NO. 43657 (12/02/1396).

\end{document}